\documentclass[final,authoryear,5p,times]{elsarticle}

\usepackage{graphicx}
\usepackage{amssymb}
\usepackage[utf8]{inputenc}
\biboptions{}

\usepackage[usenames]{color}

\journal{Astronomy \& Computing}

\begin{document}

\begin{frontmatter}



\title{The Virtual Observatory Registry}


\author[ari]{Markus Demleitner}
\ead{msdemlei@ari.uni-heidelberg.de}
\address[ari]{Unversität Heidelberg,
Astronomisches Rechen-Institut, M\"onchhofstra\ss e 12-14, 69120
Heidelberg, Germany}
\author[stsci]{Gretchen Greene}
\address[stsci]{Space Telescope Science Institute, 3700 San Martin Dr,
Baltimore, MD 21218, USA}
\author[obspm]{Pierre Le Sidaner}
\address[obspm]{VOParis/Observatoire de Paris, 61 Av de l'Observatoire
74014 Paris, France}
\author[ncsa]{Raymond L. Plante}
\address[ncsa]{National Center for Supercomputing Applications,
  University of Illinois, 1205 W. Clark St. Urbana, IL 61821}

\begin{abstract}
In the Virtual Observatory (VO), the Registry provides the mechanism with which
users and applications discover and select resources -- typically, data
and services -- that are relevant for a particular scientific problem.
Even though the VO adopted technologies in particular from the
bibliographic community where available, building the Registry system
involved a major standardisation effort, involving about a dozen
interdependent standard texts. This paper discusses the server-side
aspects of the standards and their application, as regards the
functional components (registries), the resource records in both format
and content, the exchange of resource records between registries
(harvesting), as well as the creation and management of the identifiers
used in the system based on the notion of authorities.  Registry
record authors, registry operators or even advanced users thus receive a
big picture serving as a guideline through the body of relevant standard
texts.  To complete this picture, we also mention common usage patterns and
open issues as appropriate.
\end{abstract}


\end{frontmatter}


\section{Introduction}
\label{sec:intro}

The Virtual Observatory (VO) is a distributed system -- by design, there
is no central node either running services, delivering data, or even
just a single link list-style directory.  In order to still maintain the
appearance of a single, integrated information system, users
and clients must have a means of discovering metadata of VO-compliant
resources (in the sense discussed in section~\ref{sect:recs}).  This
means is provided by the VO Registry\footnote{Written in upper case in the
following, the term ``Registry'' refers to the entire system, as opposed to the
lower-case ``registry,'' which denotes a concrete service.}.

Following the VO philosophy, the VO Registry is not a single, central
system but rather a network of several types of services, some of which
host and publish metadata collections,  while others provide capabilities
for querying such collections.  All follow standard protocols for
exchanging information between them and between them and client
software.

The VO Registry is goverened by a fairly large
set of standards; one of the goals of this paper is to review this body
of text and discuss how each standard fits into the architecture.
Anticipating some terms that will be explained later,
let us collect and arrange the relevant standards already in the
introduction\footnote{For an even bigger picture of the VO and its
components, see \citet{note:VOARCH}.}.  Where the standards have short
names in common use in the VO community, we introduce these here and
refer to the standards by their mnemonic names in the following.

\begin{itemize}
\item \emph{IVOA Identifiers} \citep{std:VOID} lays out how resources 
and resource records
in the VO are referenced.

\item \emph{Resource Metadata for the Virtual Observatory}
\citep[\emph{RM} for short;][]{std:RM}
specifies what entities need descriptions in the VO and what pieces 
of metadata these should contain to satisfy the VO's use cases.

\item \emph{VOResource}
\citep{std:VOR} lays out the basics of encoding resource metadata
information as specified in \emph{RM} in XML
and defines the basic types.  When we talk about \emph{VOResource} in
the following, we usually mean not only \citep{std:VOR} but also the
registry extensions introduced next.

\item Several \emph{Registry extensions} apply the building
blocks from \emph{VOResource} to more specialised types of services 
or interfaces.  All of these combine a definition of the metadata as
well as its XML serialisation.

\begin{itemize}
\item \emph{VODataService} \citep{std:VODS11} defines extra metadata
to describe data collections and services exposing them; in particular,
this concerns table and column metadata as well as metadata on service
parameters.

\item \emph{SimpleDALRegExt} \citep{std:DALREGEXT} defines what extra
metadata applies to services implementing several ``simple'' protocols
of the VO's Data Access Layer (DAL).

\item \emph{TAPRegExt} \citep{std:TAPREGEXT}  defines what extra
metadata applies to services implementing the Table Access Protocol TAP.

\item \emph{StandardsRegExt} \citep{std:STDREGEXT} contains resource types 
for standard texts and thus defines how standards can be referenced,
e.g., when declaring protocol support. 
\end{itemize}

\item \emph{Registry Interfaces} \citep{std:RI1} specifies how 
registries exchange the XML records defined in \emph{VOResource} and
extensions.  It also contains a Registry extension for the services
implementing Registry services themselves.  Furthermore, its current
version also defines two APIs for registry clients; in a forthcoming
version, these APIs will be dropped.

\item \emph{Registry Interfaces} re-uses the non-VO \emph{OAI-PMH}
\citep{std:OAIPMH} standard.  This Protocol for Metadata Harvesting
defined by the Open Archives Initiative governs the interactions of the
registries among themselves.
Its use by the VO is subject to several idiosyncrasies laid
out in \emph{Registry Interfaces}.

\item \emph{RegTAP} \citep{std:RegTAP} defines how registry users can query the
Registry's data content using IVOA's Table Access Protocol.  An
alternative, parameter-based API is currently being designed.  We defer
the discussion of the client APIs to a forthcoming article.
\end{itemize}

In the remainder of this paper, we will first delineate the Registry's
role in the VO and outline its scope (sect.~\ref{sect:scope}), before
establishing some basic notions on the relation between resources
and their descriptions as the VO treats it in section~\ref{sect:recs}.
Having thus introduced the concept of a resource record, in
section~\ref{sect:registries} we proceed to discuss how registries
maintain collections of them.  Section~\ref{sect:harvesting} explains
the process of transmission and dissemination of the records and the
separation of responsibilities in this process, as well as a common
implementation error that has long plagued the Registry.  The VO's way to
generate globally unique identifiers as required 
by the harvesting protocol is then
considered in section~\ref{sect:auths}.

With the basic architecture described, we proceed to discuss
current Registry content in section~\ref{sect:vorr}, in particular as
regards what resource records are contained.  This provides some insight in the
data model underlying the Registry.  For the most relevant case where
the resources described are services, special care must be taken in the
description of ``capabilities'', i.e., facilities that operate on a
client's behalf.  We give an overview of these capabilities in
section~\ref{sect:caps}.  Finally, we briefly touch the issue of the
validation of services and their descriptions in
section~\ref{sect:validation}.

\begin{figure}
\includegraphics[width=\hsize]{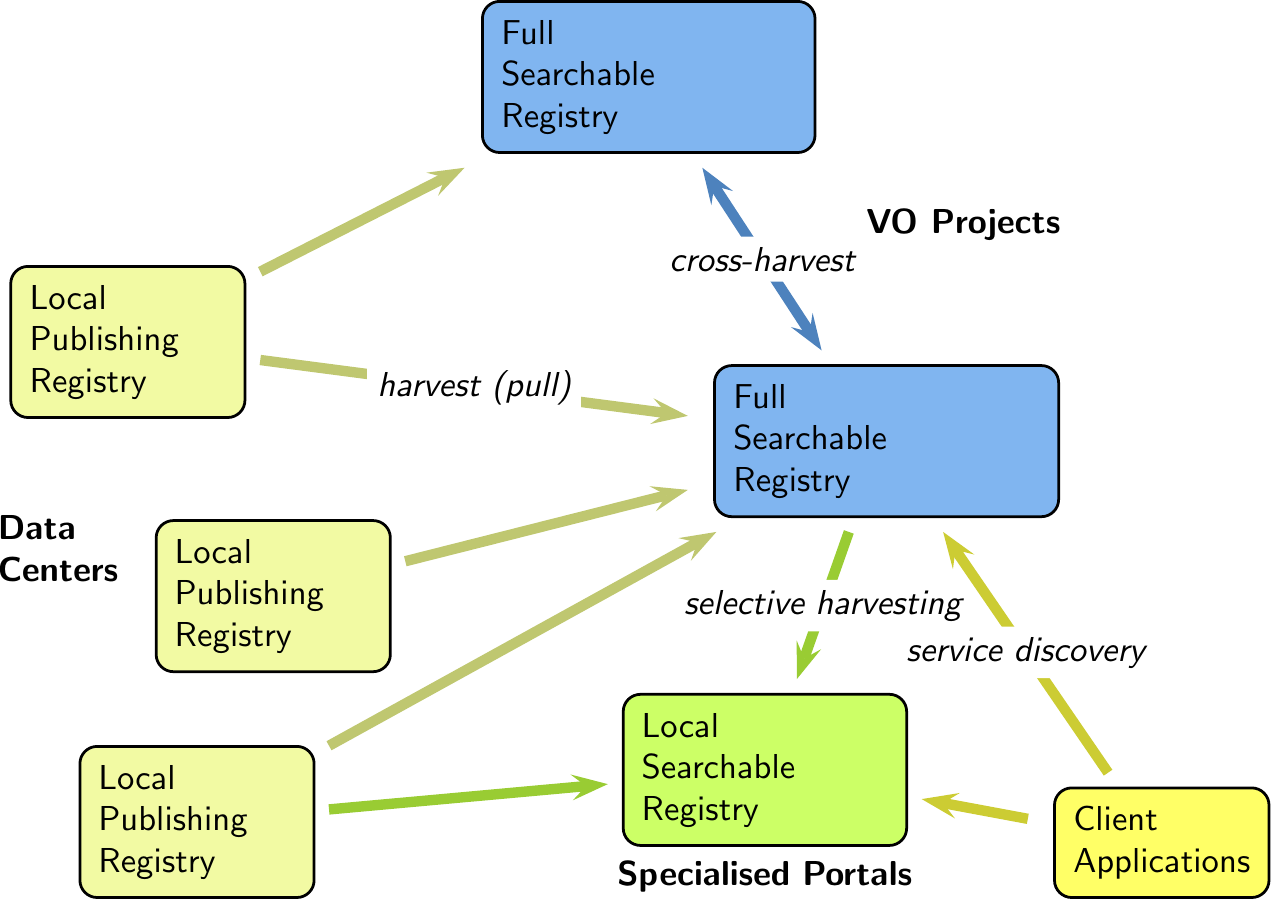}
\caption{A sketch of the registry system in the Virtual Observatory
as layed out by \citet{2007ASPC..382..445P}:
Searchable registries harvest from publishing registries operated by the data
providers.  Users and client applications can then discover VO
resources through queries to a searchable registry, either a full
searchable registry that contains everything known to the VO, or a
specialised one focused on a particular subset.}
\label{fig:arch}
\end{figure}

\section{Scope}
\label{sect:scope}

The Registry's role in the VO primarily is resource discovery.  Hence,
it must collect data sufficient to answer requests at least of the
following types (or their combination):

\begin{itemize}
\item Resources of type X (as in: image service, database service, etc)
\item Resources on topic X (defined through keywords or via a full text
search in the resource descriptions)
\item Resources with physics X (defined through waveband, observables,
queriable phenomena, etc.)
\item Resources by author(s) X
\item Resources suitable for use X
\item Resources with spatial or temporal coverage X
\end{itemize}

Once a resource record has  been located by any of these constraints, it
provides sufficient information at least to let users

\begin{itemize}
\item Assess suitability of the resource for purpose X
\item Access the resource
\item Identify who to credit for results obtained using the resource
\item Contact technical support for the resource
\end{itemize}

The VO Registry is also used to 
monitor the health and functionality of the VO.
The registries themselves are routinely validated and
curated to ensure consistency with IVOA standards, which uncovers errors
in the metadata supplied by the service operators.  Even more
importantly, services within the Registry are validated to comply to the
standards they claim to implement, and registry records, where
necessary, contain test input parameters suitable for exercising a
service.

The Registry as such is \emph{not} a mechanism of data preservation, 
and it does not provide persistent identifiers.  The identifiers
within the VO Registry, the IVORNs, are simple URIs
with an scheme of \texttt{ivo}, an authority part as discussed in
section~\ref{sect:auths}, and a local part goverened by some reasonable
restrictions on which characters are allowed to occur.  

They can be resolved to resource records and, if applicable, access URLs
by searchable registry and thus, not unlike DOIs \citep{std:DOIsystem},
introduce a level of indirection between a service identifier and its
access URLs. However, the indirection in the Registry mainly is a side
effect of the requirement to provide rich, structured metadata for the
services.  

Unlike with DOIs, an operator is free at any time to discard
identifiers, and the current VO infrastructure would stop resolving it
on a short timescale.  The conceptual reason why IVORNs as such 
are not suitable as persistent identifiers is that, as laid out in
section~\ref{sect:recs}, they are in the first place 
identifiers of the resource records.  Though the VO Registry
could be exploited as a basis for (external) data preservation services
and
persistent identifiers for resources --
\citet{2011ApSSP...1..135A} reports on one such effort --, 
it does not in itself provide such facilities.

\section{Resources and Resource Records}
\label{sect:recs}

The Virtual Observatory can be seen as a collection of \emph{resources}.
\citet{std:RM} defines a VO resource as a ``VO element that can be
described in terms of who curates or maintains it and which can be given
a name and a unique identifier.'' He goes on to name 
sky coverages, instrumental
setups, organisations, 
or data collections as examples.  In practice,
over 95\% of resources in the current VO are data services.

From the outset, it was clear that a common way of describing these
resources would be required as a very basic building block for
interoperability.  For instance, VO enabled client programs need to be able to
find out what protocols a service supports and at what ``endpoints'' --
typically, HTTP URLs -- there are available, and scientists should have
reliable and standardised ways to work out who to reference, who to
consult in case of malfunctions, and so on.  
Of course, having a standardised structure for content metadata (like
keywords, a title, description) helps writing more focused data
discovery queries as well.

Fortunately, the VO did not have to develop the technology to support
such descriptions itself, as library sciences have worked on very
comparable problems for centuries already.  The VO's registry
architecture in particular re-uses the Open Archives Initiative's
protocol for metadata harvesting \citep[OAI-PMH;][]{std:OAIPMH} for
a conceptual framework and the metadata exchange protocol, and
Dublin Core \citep{std:RFC5013} for a basis on which to build the
metadata model.

Central to \emph{OAI-PMH} is the notion of a \emph{unique identifier}, which
``unambiguously identifies an item within'' the set of resource records.
Other than that these should be URIs \citep{std:RFC2396}, \emph{OAI-PMH} does
not state details on how they should be formed.  For the VO,
\emph{IVOA Identifiers} prescribes the use of IVORNs as introduced
in section~\ref{sect:scope}.  

A somewhat subtle but nevertheless important distinction made in
\emph{OAI-PMH}
is between a resource and a \emph{resource record} containing its
description.
To see that this distinction has actual consequences, say
the data collection X contains spectra obtained using the spectrograph
S; the resource record R describes X.  Now,
during the lifetime of the instrument, S will add new data to X on every
clear night, which means the resource changes.  Nevertheless, in the
current VO R will not generally change (though it is conceivable that it will be
updated now and then, e.g., as the description might contain
rough estimates on the number of datasets contained in X).

For the converse scenario of a changing resource record with a constant
resource, suppose S is now decommissioned, while the standard defining
the content of the resource record is updated to include the spatial
coverage of the data collection.  Now, R needs an update without X
changing.

As stated above \emph{OAI-PMH} defines that its unique identifiers -- and hence
the VO's IVORNs -- always reference resource records.  As to how the
resources themselves should be referenced, \emph{OAI-PMH} declares that the
``nature of a resource identifier is outside the scope''
\citep{std:OAIPMH}. This
reservation is motivated by the library use case, where a single book
might be described by different libraries and hence have multiple
resource records. 

In the VO, it was expected that such complications would not arise as
the resource records would almost always come from the resource
publishers themselves, and no need for multiple resource records for a
single resource was foreseen.  It was therefore decided that the IVORN of
a resource record should also identify the resource itself, which
simplifies identifier generation and management significantly.

This also explains why, in \emph{OAI-PMH} messages with the metadata prefix
\texttt{ivo\_vor}
(see section~\ref{sect:registries} for details), 
the IVORN is repeated in both the header and the metadata of a
resource record.
Awareness of the distinction is relevant to registry users to
understand the meaning of the creation or update times in the resource
record (which refer to the record itself) and the dates and times given
in the curation/date child of the resource record, which pertain to the
resource.

\section{Registries}
\label{sect:registries}

Having a set of resource records alone is not enough to build a useful
system, even if they already are in a standard format.  There must also
be ways in which users can locate records of the resources relevant to
them within this set.  Therefore, systems are required enabling
service operators to feed their resource records into
the set.  Also, users must have a way to execute queries against the set.  
Both requirements
are covered by \emph{registries} within the VO.

Injection of resource records is performed by \emph{publishing
registries}.  These typically are run by service operators and deliver
resource records of a specific operator's set of services.  In addition,
for service operators who choose not to run publishing registries of
their own, both the registry at
STScI\footnote{http://vao.stsci.edu/directory/} and the registry at
ESAC\footnote{http://registry.euro-vo.org/} run publishing registries
accepting third-party resource records which are typically created using
web interfaces provided by the institutions.

Conceivably, a user looking for a resource matching some constraints
could now query each publisher's publishing registry in turn to obtain a
list of all matching VO resources.  This architecture obviously will not
scale well with the number of publishers.  It also introduces many
points of failure into the system, as all publishers would have to keep
their registries highly available to avoid a severe degradation of the
whole system.

To avoid these issues, 
retrieving resource records from the publishing registries,
joining the sets of resource records thus obtained, and offering a means
of querying this joined set to VO users is the task of specialised
agents, the \emph{searchable registries}.  The process of retrieval of
resource records from publishing registries
by a searchable registry is known as \emph{harvesting}.
To allow this harvesting, publishing and searchable registries must
agree on a common protocol.  
As mentioned in section~\ref{sect:recs}, the adoption of \emph{OAI-PMH} already
defined such a protocol for the VO.

A secondary distinction between searchable registries is between
\emph{full registries} (the term ``searchable'' is usually implied in
this case) which strive to harvest all publishing registries
in the VO and \emph{local searchable registries} which only carry a
selection of records.  An example for the second kind that is currently
in discussion is an ``educational'' registry that contains a manually
curated subset of services delivering data suitable for classroom use
(i.e., data of moderate size, with easily understood data types, etc).

The actual application of
\emph{OAI-PMH} within the VO is described in \emph{Registry Interfaces}, which in
particular defines that the VO's own resource record format is selected
in \emph{OAI-PMH} using a metadata prefix of \texttt{ivo\_vor}.  Requesting
this will make a VO-compliant registry embed \emph{VOResource} records as
discussed in section~\ref{sect:vorr} in the \emph{OAI-PMH} record's metadata
child.  VO registries
are also required to emit the much simpler Dublin core metadata
records on request and are thus interoperable with bibliographic
services outside of the VO; within the VO the much richer
\emph{VOResource} metadata is used exclusively.

One additional building block needs to be mentioned, the Registry of
Registries or RofR for short \citep{std:RofR}.  This is a special
publishing registry from which searchable registries can harvest the set
of available publishing registries to initialise or update their
internal list of registries to work on.  As such, it is a single point of
failure, as there is only one such service globally.  On the other hand,
no client code directly accesses the RofR, which means that 
an outage of the RofR does not impair
the user-visible functionality of the VO.  The main impact would be that
no new publishing registries could be added to the VO's registry system,
and existing registries' endpoints would have be be discovered from
searchable registries (which in user tools they usually are anyway).

In the current VO, the RofR also doubles as the publishing
registry for standards and other resources managed by IVOA, and it
operates a service for validating the content of publishing registries
(cf.~section~\ref{sect:validation}).

\section{Harvesting}
\label{sect:harvesting}

The VO registry system is de-centralised in both directions: A given
publishing registry does not know which searchable registries will
eventually carry its records.  An implication of this is that it cannot
notify the searchable registries when a resource record changes.  This, in
turn, implies that the searchable registries will have to poll the
publishing registries it harvests.  This is not entirely trivial, as 
the largest publishing registry in
the VO currently emits more than 100 Megabytes of resource records, and
due to paging and other delays the transfer takes about 10 minutes.

On the other hand, to keep up to date, searchable registries should poll
the publishing registries with a fairly high frequency.  Most active
searchable registries today poll once or twice a day.  To nevertheless
keep network and CPU load low, \emph{OAI-PMH} supports \emph{incremental
harvesting}.  This allows searchable registries to query publishing
registries for records updated since some point in time.

A common harvesting strategy is that searchable registries persist the
date and time of the last harvest and, on re-harvesting, query the
publishing registry for records updated since then.  Together with a
very natural-seeming (but incorrect) implementation on the part of the
publishing registry, this can lead to a loss of records with
incremental harvesting.

To see how this happens, consider a publishing registry P that, as is
usual, keeps the updated dates of its resources in a database table
to facilitate quick responses  to \emph{OAI-PMH} queries with.  The race
condition now can be exposed as follows:

\begin{enumerate}
\item A new resource record R is created at $t_1$ 
and its updated
attribute accordingly is set to $t_1$ in the record itself.  For one
reason or another, the program that ingests the updated dates for the
record into the database table does not run immediately.

\item At $t_2>t_1$, a searchable registry S harvests P and memorizes $t_2$ as
the date of the last harvest. As the database table does not contain R
yet, R is not harvested.

\item At $t_3>t_2$, the program ingesting R into the database table
is finally run, but the timestamp is taken from the resource record,
i.e., it is $t_1$.  

\item S comes back for an incremental harvest at
$t_4>t_3$ and asks for records updated after $t_2$.  As
$t_1<t_2$, R is not in the set of resources delivered.
\end{enumerate}

Hence, the record will be missed by S, which then
will not contain R.  An analogous problem exists for updates and
deletions of records.

What might seem like a fairly exotic scenario is not uncommon at all with
current registry implementations and regularly causes user-visible
differences between the content of different registries.  Some
mitigation is possible if harvesters use the time of the last-but-one
harvest to constrain their incremental queries.   
The correct solution, though, is that publishing registries set the
ingestion time as updated timestamp for their records.  As the condition
outlined above is the result of a straightforward implementation,
however, we believe in the medium term a more robust method for
incremental harvesting, presumably based on monotonously incrementing IDs,
should be put in place within \emph{OAI-PMH}, 
to make the straightforward implementation also
a race-free one.

Another sometimes misunderstood feature has to do with 
\emph{sets}.  These are
a feature of \emph{OAI-PMH} that lets archive operators define subsets of their
data holdings sharing some property.  The VO's registry interface
standard defines one such set that must be supported by all registries,
\texttt{ivo\_managed}.  This set is defined to comprise all records that
originate from the registry \emph{and} should be visible in a searchable
VO registry.  The idea behind this is that a harvesting registry can
constrain its queries to \texttt{ivo\_managed} and will not see records
from other registries even for registries harvesting other registries.
Note that set membership is a property of a registry, not of a record,
so information on set membership is lost at harvesting time.

\section{Authorities}
\label{sect:auths}

When, as in the VO, the creation of identifiers is distributed, there
needs to be a mechanism ensuring uniqueness, which in the case of the VO
Registry means making sure that no identifier is assigned to two
different resources.  In the VO, this mechanism is founded on the notion
of \emph{authorities}, which are entities creating IVORNs.  As such,
they are akin to DOI's prefixes.

As with DOI registrants owning prefixes, each
authority is assigned a namespace, within which the authority is free to
create new names, as long as some basic syntactic rules are followed.
Full identifiers are then a combination of the
authority identifier and the local part.  As long as the IVOA makes sure
authority identifiers are unique and each authority ensure uniqueness
\emph{within their namespace}, the system yields globally unique
identifiers.

Technically, authority identifiers are IVORNs (as introduced in
section~\ref{sect:scope}) that just consist of the
scheme and the URI authority part, for instance, \texttt{ivo://ivoa.net}.
By \emph{Registry Interfaces}, this must already be a
valid IVORN, i.e., refer to a resource record, which in this case must
be of the type \texttt{vg:Auth\-o\-ri\-ty}.  Resource records of this type
(``authority records'' in the following) are an ``assertion of control
over a namespace represented by an authority identifier''
\citep{std:RI1}.  In practice, the metadata
should describe
what organisational detail suggests the creation of a new authority.  In
consequence, the contact would be the person responsible for ensuring
the uniqueness of the local parts.

In addition to the usual \emph{VOResource} pieces of metadata --
discussed in detail in section~\ref{sect:vorr} --
authority records have exactly one
\texttt{managingOrg}.  This is the organisation that is responsible for
an authority, and the distinction from the authority itself is somewhat
subtle and best illustrated by an example: An observatory with an
infrared unit and an ultraviolet unit that want to avoid having to
negotiate before minting identifiers could claim the authorities
\texttt{infrared.sample}, \texttt{ultraviolet.sample}, and
\texttt{sample}.  The observatory itself would then be
\texttt{ivo://sample/org}, and it would be the managing organisation for
all the authorities.  All authority records would also list ``The sample
observatory'' (or similar) as their publisher.

Note that URI authorities are opaque and unstructured, which means that
clients are not supposed to infer any relationship from the fact that
\texttt{sample} is contained in \texttt{infrared.sample}.  There has
been a recommendation to re-use DNS names as authority IDs, which has
been largely ignored, probably because it tends to make IVORNs 
unnecessarily long.  Today, we would suggest to base authority names on
the names of national VO projects where available.

In \emph{Registry Interfaces}, the burden of ensuring the uniqueness 
of the authority names is put on the publishing registries: ``Before
the publishing registry commits the [authority] record for export, it
must first search a full registry to determine if a vg:Authority with
this identifier already exists; if it does, the publishing of the new
vg:Authority record must fail.''  Given the delays involved in
harvesting, this procedure obviously has very real issues with race
conditions, and to our knowledge, no engine for publishing registries
implements such a check.

Compared to creating and operating DOI registrants, the creation and
operation of VO authorities is thus simple, cheap and quick.  The
downside of this is that plain IVORNs do not work as persistent
identifiers as laid out in section~\ref{sect:scope}.

The construction also implies that 
only one registry is accepted as the source for registry records under
the authority (but a given registry can manage multiple authorities).
Full registries can use this mapping from authorities to their managing
registries to decide whether to ingest records they harvest when
harvesting full registries either complementary to evaluating
\texttt{ivo\_managed} or instead of it, which has in the history of the
VO Registry at times been more stable.

While name
clashes in authorities at the time they are created have not been a
problem in practice, it has frequently happened that
as authorities sometimes move from
one registry to another, the
releasing registry failed to drop its declaration of managing the
departing authority, or did not update the record's modification date,
which meant that incremental harvests would miss the update.  
This then means that two or more registries claim to managed a single
authority, which introduces
severe inconsistencies in the Registry, in particular as regards the
continual resurrection of ``zombie'' records long deleted at the
registry rightfully managing the authority.  

At this point we
believe the way to ensure a bijective mapping between authorities and
their managing registries is its manual curation at the RofR, as the
updated resource record from the accepting registry comes in and the
conflicting claims of authority can be diagnosed.

\section{VO resource records}
\label{sect:vorr}

Following \emph{RM}, VO resource records contain a fairly
comprehensive set of metadata.  All resource records must have a title,
an identifier, and a status as well as information on its content and
the curation.  They also have timestamps for the creation and the last
update of the resource record.  Additional optional metadata includes a
short name (primarily for use in cramped displays) and validation
information (cf.~section~\ref{sect:validation}).

Content metadata consists of subjects -- keywords which are supposed to
be drawn from the IVOA thesaurus \citep{std:IVOAT} --, a human readable
description, the URL of a reference page giving more information about
the resource (the \emph{reference URL}), as well as optionally
a bibliographic source -- this is what should be referenced if the data
is used -- and some additional ancillary information.
Content also allows defining relationships to other resources, examples
for which include ``mirror-of'' or ``service-for'', which is
particularly interesting for data collections to declare services
allowing access to them.

Curation metadata gives a simple provenance of a
resource: Who has created it -- to a first approximation, this usually
is the ``authors'' --, who has published it, who can repair it.  Curation
also lets publishers specify dates relevant to the history of the
resource itself (as opposed to the resource record), as, for instance,
major data additions, schema changes, or the application of corrections
for errata.

Resource records also have types, and certain types have additional
metadata.  As can be seen from table~\ref{tab:typedist}, the
overwhelming majority of resources in the current VO registry are of
type \texttt{vs:CatalogService}\footnote{Following widespread practice, we
abbreviate the namespaces \emph{VOResource} types come from with their
``canonical'' prefixes.  A review of this, including a translation from
prefixes to their namespaces, is given in section 4 of
\citet{std:RegTAP}.}.  These are access services for entities with
sky coordinates, and most VO-compliant catalog, image, or spectral
services will use this
type.

\begin{table}
\begin{center}
\begin{tabular}{|l|r|}
\hline
  \multicolumn{1}{|c|}{\vrule width 0pt height 10pt depth 5pt res\_type} &
  \multicolumn{1}{c|}{$N$} \\
\hline
  \vrule width 0pt height 10pt vs:CatalogService & 13706\\
  vs:DataCollection & 144\\
  vg:Authority & 131\\
  vr:Organisation & 76\\
  vr:Service & 48\\
  vs:DataService & 29\\
  vg:Registry & 24\\
  vstd:Standard & 7\\
  vstd:ServiceStandard & 4\\
  other & 153\\
\hline\end{tabular}
\end{center}
\caption{Distribution of resource types in April 2014, as obtained by
the prototype implementation of \emph{RegTAP} operated for GAVO in
Heidelberg (see http://dc.g-vo.org/browse/rr/q).  The XML prefixes
are as in section 4 of \citet{std:RegTAP}.
Other includes deprecated or experimental types.}
\label{tab:typedist}
\end{table}

In addition to basic \emph{VOResource} metadata, catalog services can contain
additional information on the facility and the instrument that produced
the data, whether the data is public or proprietary, on the area
covered by the data contained on the sky, and on the structure of the
table that feeds the service.  Catalog services share this metadata with
\texttt{vs:DataCollection}.

In contrast to data collections, however, catalog services have
capability metadata, which in particular lets clients work out what
protocols are available at what network endpoints. Note that capability
types and resource types are largely decoupled, and no rules are
enforced as to what resource types are allowed for which capabilities if
a resource type allows capabilities at all.
As capabilities are a fairly complex part of \emph{VOResource}, we defer their
closer discussion to section~\ref{sect:caps}.

A \texttt{vs:DataService} record is like \texttt{vs:CatalogService}, but without claiming
to be based on some tabular structure.  In retrospect, it seems doubtful
that this distinction should be reflected in the resource type, as
witnessed by its low and inconsistent use.

The interplay between \texttt{vg:Authority}, \texttt{vr:Organization},
and \texttt{vg:Registry} was
discussed in section~\ref{sect:auths}, and \emph{VOResource} just follows the
roles laid out there: \texttt{vg:Authority} in addition to the basic metadata
just gives the organisation that manages the authority,
\texttt{vr:Organisation}
allows the specification of the organisation's facilities and
instruments, and \texttt{vg:Registry} lists the authorities it manages, whether
it is a full registry, and it has capabilities.  Whether a registry is
searchable or publishing or both is determined by its capabilities in
\emph{Registry Interfaces}.  In \emph{RegTAP}, data model identifiers
from \emph{TAPRegExt} are used for registry API discovery instead.

While few in number, records of types \texttt{vstd:Standard} and
\texttt{vstd:ServiceStandard} are nevertheless important.  They serve as
destinations for references to standards as required in, e.g.,
capability records as discussed below.  Such records allow the
declaration of the various versions of a standard, associated XML
namespace URIs, and also the declaration of terms.  This latter feature
provides a relatively lightweight way to generate IVORNs for certain
concepts standards might need.  In the registry extension for TAP
\citep{std:TAPREGEXT}, for example, this mechanism is used to introduce
identifiers for output formats not distinguishable by MIME type.
Service standard records, in addition, allow a simple specification of a
standard service's interface.

We finally mention the status attribute of \emph{VOResource} records.
It is distinct from but related to \emph{OAI-PMH}'s \texttt{status} element
optionally present in OAI headers; there, \texttt{status} take the
single value \texttt{deleted}, which should cause a harvesting registry
to remove a resource record with the same identifier it may have stored
from previous harvests (provided it uses the \texttt{ivo\_vor} metadata
prefix consistently). As VOResource describes the resource rather than
the resource record, its \texttt{status} attribute in addition can
assume the values \texttt{active} (which for resource records is implied
by the fact they can be harvested) and \texttt{inactive}. This latter
value is intended as a measure for publishers of third-party resource
records when they suspect a resource registred through them has gone
unmaintained but do not want to remove the resource record entriely.  It
is a feature rarely used, and the upcoming Registry APIs do not expose
inactive resources to clients, since to them nonresponsive services
coming back from registries are an annoyance regardless of prospects for
the service's restoration.

\section{Capabilities}
\label{sect:caps}

Resource types that offer endpoints for interaction (services,
registry) also contain zero or more capability elements.  Capabilities
essentially are \emph{VOResource}'s way to describe the possible interactions
with a resource.\footnote{An exception to the interact-through-capability
concept is \texttt{vs:DataCollection}'s accessURL, which allows retrieval of the
data and is a top-level attribute of the resource.}

\emph{VOResource}'s basic capability element consists of optional validation
information, and optional human-readable description, and zero or more
interfaces.

The interfaces are again typed, with most interfaces in the current VO
being one of \texttt{vs:ParamHTTP} -- an interface for operation by HTTP and HTTP
request parameters (about 64\%) -- and \texttt{vr:WebBrowser} -- services based
on HTML forms (about 35\%).  The remaining interfaces are a few
SOAP-based services, the special OAIHTTP type used by publishing
registries, and some types from abandoned standards.

Interfaces have one or more access URLs, where we expect that the next
version of \emph{VOResource} will restrict this to exactly one.  In addition,
a role attribute should be set to \texttt{std} if the interface is a standard
interface for the standard the capability claims to implement.  In that
case, a version attribute can give the version of this standard.  In current VO
practice, this version attribute is typically ignored, as incompatible
standards are told apart by the standard identifier of the capability.

Derivations of \texttt{vr:Interface} may have additional properties.  In
particular, \texttt{vs:ParamHTTP} declares a result type -- supposed to be a MIME
type -- and the input parameters with their names, UCDs, and types,
expressed in a simplified type system.  This is a cross-protocol way of
discovering the parameter metadata which should be provided in addition
to protocol-specific means.  Compared to the parameter declarations
emitted from metadata queries in the VO's image and spectral access
protocols SIAP and SSAP
\citep{std:SIAP,std:SSAP}, parameter declarations in interfaces are less
expressive, since the VOTable PARAMs employed in SIAP/SSAP metadata can have
VALUES children giving ranges or possible values for enumerated
parameters.  It is somewhat unfortunate that the same kind of information
is exposed in two non-equivalent ways.

In addition to these basic capability metadata, registry extensions can
define capabilities with richer metadata.  For instance,
\emph{SimpleDALRegExt} defines things
like test queries, limits to search and response sizes, but also the
kind of data contained, which for the image access protocol SIAP 
declares whether the service
returns cutouts, pointed observations, mosaiced images, or is an
atlas-type service.  The most complex capability structure so far is the
one for the Table Access Protocol TAP (\emph{TAPRegExt}),
which exposes many aspects of
the TAP service and the languages supported by it.  In the context of a
paper on the registry, TAPRegExt's 
\texttt{dataModel} element deserves particular
attention.  It contains an IVORN of a standard defining a data model,
more specifically a set of relational tables.  This can be used to
locate TAP services having these tables.  Both Obscore -- a table schema
for observational products, \citet{std:OBSCORE} -- and the upcoming
\emph{RegTAP} standard use this mechanism to enable service discovery.

Capabilities are not only used directly in the registry.
The VOSI and DALI standards \citep{std:VOSI, std:DALI} mandate that
services should also emit the capability elements on a specialised
endpoint next to the science endpoints.  An example for where these
endpoints are already in everyday use is again TAP, where clients determine
the details of a TAP service (user defined functions, support for
optional features, output formats, limits, etc) without having to
consult a registry.

\section{Validation}
\label{sect:validation}

In a distributed system in which many parties operate services, partly
using custom implementations, it is inevitable that not all services
actually comply to the standards they claim to implement.  With a
complex system like the VO Registry, it is not trivial to even write
correct and complete resource records, let alone follow all rules
ensuring that a publishing registry fits into the whole system.  Hence,
validation on many levels is crucial for maintaining the
integrity of the VO.

As regards the \emph{VOResource} records themselves, their validity essentially
is equivalent to their compliance to the XML schema files that accompany
the pertinent standards.  For a publishing registry, a large number of
further properties need to be checked, for instance a correct
implementation of \emph{OAI-PMH}, the definition of the authorities managed by
the registry, the support of the \texttt{ivo\_managed} set, and so
forth.

A service performing such a validation is operated at the RofR, and it
has proven instrumental for building a working Registry system.  In
particular, publishing registries that try to enlist themselves in the
RofR are validated and can only enter if they are valid.

Registries may become non-compliant after this initial validation due to
software updates or, more commonly, invalid registry records entering
the set of resource records.  No automatic re-validation is taking
place, and registries that become invalid are not removed from the RofR.
Relying on the registry operators to re-validate and repair their
services has so far proven sufficient for keeping the VO Registry
operational.

There is, however, a second and much larger aspect to validation:
resource validation.  This is another case in which the distinction
between resource record and the resource itself becomes relevant -- a
valid resource record might very well describe a service that does not
comply to the underlying standard.  Validating a resource means
examining as many aspects of its operation as possible.  While this
validation can in principle be performed by anyone, a publishing
registry is a natural place for the operation of a service validator:
(a) it already has the metadata available; (b) it has a means to
disseminate its results.

As to (a), this metadata obviously includes the access URL and the
standard implemented.  However, meaningful validation typically requires
additional metadata, in particular parameters that must return a
non-empty response.  \emph{SimpleDALRegExt} contains elements designed for
that purpose.  For instance, the cone search capability has a
\texttt{testQuery} element that separately lists values for the \texttt{RA},
\texttt{DEC}, and \texttt{SR} parameters that VO cone searches require.
In actual use, it turned out that
separating out the individual parameters of protocols did not
significantly help either validators or other VO components.  In the
most recent simple DAL extension, the one for SSAP, \texttt{testQuery}
hence admits the specification of a complete query string otherwise
opaque to the validator.

As to (b), \emph{VOResource} introduces a validation type that allows
operators of validators to communicate their results.  It consists of a
numeric code from \emph{RM} 
and a mandatory URI identifying the validating entity.  The
numeric code currently ranges between 0 -- ``has a description that is
stored in a registry'' -- and 4 -- ``meets additional quality criteria
set by the human inspector,'' where from 2 up there is a requirement
that the resource described exists and has been ``demonstrated to be
functionally compliant.''

A resource record may contain validation information for both the full
record and for a single capability.  While the exact semantics of this
distinction is not easy to define, the rough guideline from
\emph{VOResource} suffices for a useful interpretation.  According to
this, when a validation level is given for a resource, the
``grade applies to the core set of metadata,'' whereas ``capability and
interface metadata, as well as the compliance of the service with the
interface standard, is rated by validationLevel tag in the capability
element.''

Validation information is different from the rest of the resource record
in that it is the only part designed to be changed by a third party on
the way from the resource record author through publishing and
searchable registry to the resource record consumer.  It is also the
only piece of information that a harvester should accept from a resource
record it harvests from somewhere other than the originating
registry.

As almost all other aspects of the VO, validation is distributed.
Conceptually, everyone is free to offer a harvesteable registry handing
out validity assessments.  In actual experience, validity assessments
actually differ between various validating entities, for example because
the feature sets exercised by the various validators are different.
Several organisations in the VO operate validators, for instance, the
Observatoire de Paris \citep{voparisvalidator}, which also keeps
a history of the performance of services such that it is easy to
diagnose services that have been unresponsive or severely degraded for
extended periods of time.

\section{Conclusions}
\label{sec:conc}

The Registry is the Virtual Observatory's answer to the need for
structured, global, and detailed resource discovery.  It exposes
to clients a wealth of metadata while not
introducing a single point of failure.  This is enabled by a strictly
defined metadata format, the use of standard protocols in the
communication between registries, judicious use of cross-harvesting,
authority management, and continuous validation.

The article reviews how a set of standards by both the IVOA and external
communities lay the foundations for the whole Registry system consisting
of (cf.~Fig.~\ref{fig:arch} for a graphical representation of this):

\begin{itemize}
\item publishing registries run by the providers of the science
services (or on their behalf) that inject the resource records in a
flexible and extensible metadata format,
\item searchable registries that harvest the publishing registries (and
potentially each other)
\item a single registry of registries facilitating the initial discovery
of registries (but is not important in daily operation of the Registry,
as its content is also available from all full registries),
\item and user interfaces and APIs provided by the searchable registries
exposing the Registry contents to queries and inspection (these will be
discussed in a forthcoming article).
\end{itemize}

The Registry has additional roles to play on top of resource discovery.
For example, information on the publishers, creators, and maintainers
of the resources are available in a standardised way.  This lets
client software present the VO user with information on who to credit
in a study using data obtained from registred services, 
or to find out where to direct questions in
case of technical malfunction or scientific issues.

The success of the development of a resource, and in particular service,
registry within the VO
may also be seen from the adoption of the underlying
technologies in similar projects in other fields, for instance a VO-like
effort in molecular and atomic spectroscopy called VAMDC
\citep{2011ASPC..442...89W}.

\section*{Acknowledgements}

This work was in part supported by the German Astrophyiscal Virtual
Observatory GAVO, BMBF grant 05A11VH3.

\section*{References}

\bibliographystyle{elsarticle-harv}
\bibliography{registry}

\end{document}